\documentclass[10pt,conference]{IEEEtran}               

\usepackage{graphicx}          
                               
\usepackage{amsfonts,amsmath,amssymb}
\usepackage{algorithmicx,algpseudocode}
\usepackage{algorithm}

\newtheorem{definition}{Definition}
\newtheorem{remark}{Remark}

\newtheorem{problem}{Problem}
\newtheorem{proof}{Proof}
\newtheorem{proposition}{Proposition}

\makeatletter
\def\endthebibliography{%
  \def\@noitemerr{\@latex@warning{Empty `thebibliography' environment}}%
  \endlist
}
\makeatother

\IEEEoverridecommandlockouts                              
\title{\LARGE \bf
Non-uniform Sampled Motion Planning for Continuous-time STL
}

\author{Guang Yang$^{1}$, Calin Belta$^{2}$ and Roberto Tron$^{2}$
\thanks{$^{1}$Guang Yang is with Division of Systems Engineering at Boston University, Boston, MA 02215 USA. Email: {\tt\small gyang101@bu.edu}}
\thanks{$^{2}$Roberto Tron,$^{2}$Calin Belta are with the Department of Mechanical Engineering at Boston University, Boston, MA 02215 USA. Email: {\tt\small tron@bu.edu, cbelta@bu.edu}}
}

\begin{document}

\maketitle

\begin{abstract}                          
This paper presents an offline motion planner for linear cyber-physical systems that satisfy a continuous-time Signal Temporal Logic (STL) specification, in which controls are applied in a Zeroth-order Hold (ZOH) manner. The motion planning problem is formulated as a Mixed-integer Program (MIP) with nonuniform control updates.  We develop a novel method to obtain bounds of Control Barrier Functions (CBF) and linear predicates to render both spatial and temporal requirements. The theoretical results are validated in numerical examples.
\end{abstract}

\section{Introduction}
\subsection{Motivation}
The autonomous applications that involve cyber-physical systems have gained significant popularity in recent years, such as a self-driving car, perform search-and-rescue with aerial drones, and assistant robots for home or medical applications. Within these applications, motion planning is a crucial component to ensure the assigned tasks can be executed correctly and safely. There are two significant challenges to perform motion planning on these systems: First, the physical systems exist in a world in continuous time, but has to be controlled and sampled discretely with digital computers. Second, the applications often involve complex mission specifications that include temporal and spatial constraints. For example, a search and rescue mission with a drone could have the following mission specification: the drone has to eventually visit regions with deadlines, while always avoiding unsafe areas. Besides, it has to maintain its velocity within the desired range. In summary, both spatial and temporal constraints could appear in complex missions that involve autonomous systems. In this paper, we attempt to address these challenges by creating a safe and efficient motion planner.

\subsection{Problem Overview}
 We consider a trajectory planning problem for a continuous-time linear system with non-uniform samplings and control updates. In this paper, a motion planning problem for a continuous-time linear system is considered. The objective is to steer a system trajectory to satisfy a continuous-time STL by formulating and solving a Mixed-integer Program (MIP).

 \subsection{Approach Overview and Contribution}
 In our proposed method, we formulate a MIP with constraints obtained from a Signal Temporal Logic (STL) specification in such a way that continuous-time satisfaction is guaranteed with discrete Zeroth-order Hold (ZOH) control updates.  The linear STL predicates are encoded via their robustness functions into sets of linear constraints, guaranteeing satisfaction on a finite discrete set of time instants. The temporal operators, such as Always (\textbf{G}) and Eventually (\textbf{F}), can be interpret as \textit{set invariance} and \textit{finite time reachability}, respectively. To render \textit{set invariance}, we use CBFs to derive constraints that guarantee the state trajectory stays within a set that is defined by the predicate for a fixed time interval. To achieve \textit{finite time reachability}, we use the lower bound of a predicate that guarantees the state trajectory reaches the desired set in a finite time. The major contributions are listed as follows:
\begin{itemize}
  \item We propose a novel encoding method for Eventually operator (\textbf{F}) in the STL by using lower bounds for predicates in time. 
  \item We introduce a heuristic method to determine control update and sampling instants. 
\end{itemize}

\subsection{Related Work}
Temporal Logic (TL)-based control has been widely used in the context of persistent surveillance \cite{leahy2016persistent}, traffic control \cite{sadraddini2016model} and distributed sensing \cite{serlin2018distributed}. While originating from the field of formal methods \cite{baier2008principles}, TLs are now used to describe specifications for a variety of system behaviors, as attested by the proliferation of many different specialized languages (such as Linear Temporal Logic\cite{pnueli1977temporal}, Computation Tree Logic\cite{clarke1981design} and Time Window Temporal Logic \cite{VASILE201727}). For applications that require the definition of real values with bounded time constraints,  Signal Temporal Logic (STL) \cite{maler2004monitoring} and Metric Temporal Logic (MTL) \cite{koymans1990specifying} have been introduced.

The notion of STL robustness over real-valued signals \cite{fainekos2009robustness}, also known as space robustness, provides a quantitative semantics of how well a signal satisfies a given STL formula. There are other ways to measure STL robustness, such as average based robustness \cite{mehdipour2019arithmetic} that use arithmetic geometric mean to account for the frequency of satisfactions of a given STL. In this paper, we only consider space robustness as a measure of satisfaction. In discrete-time, it is possible to encode the robustness function of a formula into the constraints of a Mixed Integer Program (MIP), thus allowing for relatively efficient control synthesis \cite{raman2014model,raman2015reactive,sadraddini2015robust}. The major drawback of this type of approach is that it is limited to the discrete-time setting for verifying the satisfaction of each predicate (since all time instants need to be represented with variables in the MIP); if the same paradigm is applied to the discretization of continuous-time systems, it does not guarantee the satisfaction of the formula in between two sampled time steps. Moreover, on the one hand, practical systems evolve in continuous time, and time intervals in the specification can also involve arbitrary (application-driven) continuous-time intervals. On the other hand, there is usually a limitation on the control and actuation rates that can be achieved, and the control updates cannot be generally assumed to coincide with the time intervals in the specification. 

Control Barrier Functions (CBFs, first introduced in \cite{wieland2007constructive}), are related to Control Lyapunov Functions (CLF), but instead of stability, they guarantee that the trajectories of a system remain in a pre-defined \emph{forward invariant} set. CBFs have been extended to Exponential CBFs \cite{nguyen2016exponential} and High Order CBF (HOCBF) \cite{Xiao2019} for systems with a relative degree higher than one. CBFs have been applied to adaptive cruise control \cite{ames2014control}, swarm manipulation \cite{borrmann2015control}, heterogeneous multi-robot manipulation \cite{wang2016safety}, and bipedal robotic walking \cite{hsu2015control}. A typical CBF formulation involves a continuous-time system and results in a Quadratic Program (QP) that needs to be solved at every control update. For real-world systems with discrete-time updates, the computed controls are applied in a Zero Order Hold (ZOH) manner, but special care needs to be taken in order to ensure that the CBF constraints hold true in between the two control updates \cite{ghaffari2018safety},\cite{8814657}. 

There exist some work \cite{lindemann2019control} that combines TLs with CBFs using continuous dynamics in which the formulation predicates are guaranteed to be satisfied only at discrete times. A similar approach can be seen in \cite{garg2019control}, where a Model Predictive Control (MPC) approach is introduced to satisfy spatial and temporal constraints. Both works do not guarantee continuous-time satisfaction. We addressed the issue of rendering set invariance with Always (\textbf{G}) operator under discrete control inputs in \cite{yang2019continuous}. To the best of our knowledge, there have been no attempts to address the Eventually (\textbf{F}) operator in continuous time. 

\section{Problem Statement}
\begin{problem}
\label{problemFormulation}
Given a linear system \eqref{continousSystem} with initial state $x_0 \in X \subseteq \mathbb{R}^{n}$ and a continuous-time STL formula $\varphi$ with horizon $t_f$, synthesize a sequence of discrete control inputs $u[t_k], k=1,...,N$, that minimizes a cost function $J(u)$ over the horizon, while the trajectory satisfies the formula $\varphi$.
\end{problem}

\section{Preliminaries}
\subsection{Notation}
We use $\mathbb{Z}$ and $\mathbb{R}^n$ to denote the set of integers and the $n$-dimensional real space, respectively. A function $f: \mathbb{R}^n \mapsto \mathbb{R}^m$ is called \emph{Lipschitz continuous} on $\mathbb{R}^n$ if there exists a positive real constant $L\in \mathbb{R}^{+}$, such that $\|f(y)-f(x)\| \leq L \|y-x\|, \forall x,y \in \mathbb{R}^n$. Given a continuously differentiable function $h:\mathbb{R} \mapsto \mathbb{R}$, we use $\dot{h}$ to denote first order time derivative and $h^{(r)}$ to denote its $r$-th order derivative with respect to time $t$. A continuous function $\alpha:[-b,a) \mapsto [-\infty,\infty)$, for some $a>0, b>0$, is called a class $K$ if $\alpha$ is strictly increasing, and $\alpha(0)=0$. 

\subsection{System Dynamics}
\label{sec:dynamical-system}
Consider a continuous-time linear system:
\begin{equation}
    \dot{x}(t) = Ax(t)+Bu, \label{continousSystem}
\end{equation}
where system matrices are $A \in \mathbb{R}^{n\times n}$, $B\in \mathbb{R}^{n\times m}$, while $x\in \mathbb{R}^{n}$ and $u\in\mathbb{R}^{m}$ represent the state and control inputs.

We assume that we are only able to update the control inputs only at regular discrete sampling instants. We denote $t_k$ as the $k$-th sampling time instant, and the time interval between control updates as $\tau = t_{k+1}-t_k, k=\{1,2,\ldots\}$. For $t\in[t_k,t_{k+1})$, we implement the Zeroth-order Hold control which holds a control signal at $t_k$ constantly until $t_{k+1}$. For each update interval, the dynamics \eqref{continousSystem} can be exactly integrated as
\begin{equation}\label{eq:discretizeEquation}
  x(t) =  e^{A(t-t_k)}x[t_k] + \int_{t_k}^{t_{k+1}} e^{A(t-s)} \mathrm{d}s Bu[t_k],
\end{equation}
for $t_{k} \leq t < t_{k+1}$.
Let $V^{-1}QV$ be the Jordan decomposition of $A$, where $V$ an invertible matrix, and $Q$ a block-diagonal matrix containing $\kappa$ Jordan blocks. We denote $s(i)$ and $\lambda_i$ the size and eigenvalue associated with $i$-th Jordan block, respectively, $i\in\{1,\ldots,\kappa\}$. With this decomposition, we can rewrite \eqref{eq:discretizeEquation} as:
\begin{align} \label{eq:generalForm}
    x(t) &= e^{A(t-t_k)}x[t_k] \\\nonumber
    &+ e^{A(t-t_k)} V \int_{t_k}^{t_{k+1}} e^{-Qs}  \mathrm{d}s V^{-1} Bu[t_k].\\\nonumber
\end{align}

\subsection{Higher Order Control Barrier Function}
We define an invariant set using time-varying function $h:\mathbb{R}^n \times [t_0, \infty) \mapsto \mathbb{R}$ that is $r_b^{th}$ order differentiable in the form
\begin{align}\label{eq:rb-order-functions}
    &\Psi_0(x,t) = h(x,t), \\ \nonumber
    &\Psi_1(x,t) = \dot{\Psi}_0(x,t) + \alpha_1(\Psi_0(x,t)), \\ \nonumber
    &\dots \\ \nonumber
    &\Psi_{r_b}(x,t) = \dot{\Psi}_{r_b-1} +  \alpha_{r_b}(\Psi_{r_b-1}(x,t)). \\ \nonumber
\end{align}
To ensure the state trajectory remain within the set, we denote a series of sets with functions $\Psi_i$ as 
\begin{align}\label{eq:assosicateSets}
    &\mathfrak{C}_0 = \{x \in{\mathbb{R}^n}|\Psi_0(x)  \geq 0 \}, \\ \nonumber
    &\mathfrak{C}_1 = \{x \in{\mathbb{R}^n}|\Psi_1(x)  \geq 0 \}, \\ \nonumber
    &\dots \\ \nonumber
    &\mathfrak{C}_{r_b} = \{x \in{\mathbb{R}^n}|\Psi_{r_b}(x)  \geq 0 \}. \\ \nonumber
\end{align}
\begin{definition} \cite{Xiao2019}
Given the functions defined in \eqref{eq:rb-order-functions} and safety sets \eqref{eq:assosicateSets}, the $r_b^{th}$ order differentiable function $h:\mathbb{R}^n \times [t_0, \infty) \mapsto \mathbb{R}$ is a Higher Order Control Barrier Function (HOCBF) for system \eqref{continousSystem} if there exists class $K$ functions $\alpha_1,\dots,\alpha_{r_b}$ such that 
\begin{equation}\label{eq:hocbf}
    \Psi_{r_b}(x(t),t) \geq 0
\end{equation}
for all $(x,t) \in \mathfrak{C}_0 \cap \dots \cap \mathfrak{C}_{r_b} \times [t_0,\infty) $. The system is forward invariant.
\end{definition}

\begin{remark}
In this paper, we use Exponential Control Barrier Function (ECBF) \cite{nguyen2016exponential}, which is a special case of the HOCBF \cite{Xiao2019}. 
\end{remark}

\subsection{Signal Temporal Logic}
The syntax of STL is recursively defined as:
\begin{equation*}\label{eq:STLsyntax}
    \varphi := \top | \mu | \neg \varphi | \varphi_1 \land \varphi_2 | \varphi_1 \vee \varphi_2| \textbf{F}_{[a,b]} \varphi| \textbf{G}_{[a,b]} \varphi |\varphi_1 \mathcal{U}_{[a,b]} \varphi_2
\end{equation*}
,where $\top$ is the Boolean constant \emph{true}, and $\mu$ is a predicate.
We consider predicates $\mu_i$ of the form
\begin{equation*} \label{eq:yOut}
    \mu := h(x) \geq 0,
  \end{equation*}
where $h$ is a linear function over the states
of (\ref{continousSystem}). The \textit{Eventually} temporal operator $\textbf{F}_{[a,b]} \varphi$ specifies that $\varphi$ holds true at some time step between $[a,b]$. The \textit{Always} operator $\textbf{G}_{[a,b]} \varphi$ states that $\varphi$ must holds true $\forall t \in [a,b]$. To state that a signal $y$ satisfies a specification (formula) $\varphi$ at time $t$ we use the notation $x(t)\models \varphi$. The STL semantics is the defined as follows:
\begin{equation}
    \begin{aligned}
    &(x,t) \models \mu \Leftrightarrow h(x) \geq 0 \\
    &(x,t) \models \neg \mu \Leftrightarrow \neg ((x,t) \models \mu)\\
    &(x,t) \models \mu_1 \land \mu_2 \Leftrightarrow (x,t) \models \mu_1 \land (x,t) \models \mu_2 \\
    &(x,t) \models  \textbf{F}_{[a,b]} \mu \Leftrightarrow \exists t' \in [t+a, t+b] s.t. (x,t') \models \mu \\
    &(x,t) \models  \textbf{G}_{[a,b]} \mu \Leftrightarrow \neg \textbf{F}_{[a,b]} (\neg \mu)\\
    &(x,t) \models \varphi_1 \mathcal{U}_{[a,b]} \varphi_2 \Leftrightarrow \exists t' \in [t+a, t+b] \\&s.t. (x,t') \models \varphi_2 \land \forall t'' \in [t,t'], (x,t'')\models \varphi_1.\\
    \end{aligned}
\end{equation}
All STL temporal operators have bounded time intervals in continuous time. 
The {\em horizon} of an STL formula is the minumum time needed to decide its satisfaction.

\subsection{Mixed Integer Formulation for STL}
\label{sec:STL-mixed-integer}

In this section, we review the binary encoding of STL robustness using mixed-integer constraints proposed in \cite{sadraddini2015robust}. This encoding is based on the big-$M$ method, where a sufficiently large number $M$ is introduced to enforce logical constraints. For the $i$-th predicate $\mu_i$ and the corresponding binary variable $z_{\mu_i}[t_k] \in \{0,1\}$, we use the constraints 
\begin{align*}
    h_i(x(t_k)) \leq M z_{\mu}[t_k], &&  -h_i(x(t_k)) \leq M (1-z_{\mu}[t_k]),
\end{align*}
to establish the relation
\begin{equation*}
h_i(x(t_k)) \geq 0 \iff z_{\mu}[t_k] = 1
\end{equation*}
at time $t_k$.

For an STL formula $\varphi$ with horizon $N$, we denote $z_{\varphi}[t_k] \in \{0,1\}$, with $(x,t) \models \varphi \iff z_{\varphi}[t] = 1$. 
We also denote $z_{\varphi_i}[t]^k \in \{0,1\}$ for the $i$-th subformula which is recursively defined based on the STL semantics \eqref{eq:STLsyntax}.

Given an STL formula $\varphi$, we can recursively encode the rest of the logical operators by using the binary variables of subformula and predicates as shown in Table~\ref{tb:STL_formulation} (we dropped the $k$ for simplicity in the table). 
\begin{table}[hb] 
\centering
\begin{center}
\scalebox{0.9}{
\begin{tabular}{|c|c|c|}
\hline
&\textbf{Definition}&\textbf{Encoding Rule} \\
\hline
$\wedge$&  $z_{\varphi}[t] = \wedge_{i=1}^{p} z_{\psi_i}[t]$ & $
      z_{\varphi}[t] \leq z_{\psi_i}[t],
      z_{\varphi}[t] \geq 1-p+\sum \limits_{i=1}^{p} z_{\psi_i}[t]$\\
\hline
$\vee$ &$z_{\varphi}^t = \vee_{i=1}^{p} z_{\psi_i}[t]$&$z_{\varphi}[t] \geq z_{\psi_i}[t],
      z_{\varphi}[t] \leq \sum \limits_{i=1}^{p} z_{\psi_i}[t]$ \\
\hline
$\neg$&$z_{\varphi}[t] =  \neg z_{\psi}[t]$&$z_{\varphi}[t] = 1- z_{\psi}[t]$\\
\hline
$\textbf{F}$&$\varphi = \textbf{F}_{[a,b]} \psi$ & $z_{\varphi}[t] = \bigvee \limits_{\tau=t+a}^{t+b} z_{\psi_i}^{\tau}$\\
\hline
$\textbf{G}$ &$\varphi = \textbf{G}_{[a,b]} \psi$&$z_{\varphi}[t] = \bigwedge \limits_{\tau=t+a}^{t+b} z_{\psi_i}^{\tau}$\\
\hline
$\mathcal{U}$ & $\varphi = \psi_1 \mathcal{U}_{[a,b]}\psi_2 $ &$\mathbf{G}_{[0,a]} \psi_1 \wedge \mathbf{F}_{[a,b]}\psi_2 \wedge \mathbf{F}_{[a,a]}\psi_1 \mathcal{U}\psi_2$\\
\hline
\end{tabular}}
\label{tab2}
\end{center}
\caption{STL Encoding with Mixed-integer}\label{tb:STL_formulation}
\end{table}

\section{Method}
This section contains the main theoretical and algorithmic contributions of this paper. We introduce the notion of CBF lower bound and predicate lower bound for Always operator ($\textbf{G}$) and Eventually operator ($\textbf{F}$), respectively. We first present the CBF formulation in section \ref{sec:CBFconstraintForlinearSys}. In section \ref{sec:CBFlowerbound}, we demonstrate how to obtain the lower bound of a given linear CBF constraint using mixed-integer encoding. In section \ref{sec:CBFSTLencoding}, we show how certain STL formulas can be encoded as CBF constraints. In \ref{sec:predicate lowerbound}, we introduce the notation of predicate lower bound in continuous time. Finally, the MIP based motion planner is formally defined in section \ref{sec:miqp}.

\subsection{Predicate Sets}
To ensure our planned trajectory satisfies the continuous-time property, we encode each linear predicate as a predicate set.  Let us define a predicate set $C$
\begin{equation}\label{eq:safetySet}
C = \{x \in{\mathbb{R}^n}|h(x)  \geq 0 \},
\end{equation}
and use $\partial{C}$ 
and $Int(C)$ to denote 
the boundary and the interior of $C$. In this paper, we consider affine predicate as a smooth function in the form
\begin{equation} \label{eq:linSafetyConstr}
    h(x) = \nu^{T} x + \gamma,
\end{equation}
where $\nu \in \mathbb{R}^n$ and $\gamma \in \mathbb{R}$. We define the Lie derivative of a smooth function $h(x(t))$ along the dynamics \eqref{continousSystem} as $\pounds_{Ax} h(x) := \frac{\partial h(x(t))}{\partial x(t)} Ax(t)$, $\pounds_{B} h(x) := \frac{\partial h(x(t))}{\partial x(t)} B$. The relative degree $r_b \geq 1$ is defined as the smallest natural number such that $\pounds_{B} \pounds_{Ax}^{r_b-1} h(x) u \neq 0$. The time derivatives of $h$ can then be expressed as
\begin{align} \label{eq:rbh}
h^{(r_b)}(x) &= \pounds_{Ax}^{r_b}h(x) + \pounds_{B}\pounds_{Ax}^{r_b-1}h(x)u.
\end{align}
Given the linear system \eqref{continousSystem} and the time derivative \eqref{eq:rbh}, we can obtain 
\begin{equation} \label{eq:rbh_for_linear_system}
h^{(r_b)}(x) = \nu^T(A)^{r_b}x+\nu^T(A)^{r_b-1}Bu,
\end{equation}
where $(A)^{r_b}$ is the $r_b$-th power of $A$.

\subsection{CBFs for Linear Constraints}
\label{sec:CBFconstraintForlinearSys}
From the closed form solution for the dynamical system and predicate \eqref{eq:linSafetyConstr}, we can write the CBF constraint \eqref{eq:hocbf} at the $k$-th update instant as
\begin{multline} \label{eq:zeta_func}
\zeta_k(t)=\sigma + \sum_{i=1}^{\kappa} \sum_{j=0}^{s(i)-1} c_{k,i,j}^{(x)T} x[t_k] e^{\lambda_i t} t^j+{c_{k,i,j}^{(u)T}}u[t_k] e^{\lambda_i t}t^j\\ \zeta_k(t)\geq 0, \forall t \in [t_k,t_{k+1}].
\end{multline}
where $c_{k,i,j}^{(x)} \in \mathbb{R}^{n}$, $c_{k,i,j}^{(u)} \in \mathbb{R}^{n}$, and $\sigma\in\mathbb{R}$ are constants obtained by solving the matrix exponentials in \eqref{eq:generalForm} and carrying out the subsequent matrix-vector calculations. 

From a computational standpoint, the main difficulty in enforcing \eqref{eq:zeta_func} is the fact that inequality needs to hold on an entire interval of $\tau$. An equivalent constraint could be obtained by taking the minimum of $\zeta(t)$ over the same interval, and then enforcing the inequality on this minimum. However, analytically computing such a minimum is not trivial. To sidestep this difficulty, we decompose the sum in \eqref{eq:zeta_func} into the following terms:
\begin{align}
&\zeta_{k,i,j}^{(x)}(t) =  c_{k,i,j}^{(x)T} x[t_k] e^{\lambda_i t} t^j,\\ \nonumber
  &\zeta_{k,i,j}^{(u)}(t) =  {c_{k,i,j}^{(u)T}}u[t_k] e^{\lambda_i t} t^j,\\ \nonumber
  &i=1,\ldots,\kappa,\\ \nonumber
  &j=0,\ldots,s(i)-1,
\end{align}
and  we introduce a set of slack variables $\beta^{(x)}_{ij}$, $\beta^{(x)}_{ij}$, $i = 1,\ldots,\kappa, j=0,\ldots,s(i)-1$ such that
\begin{equation}
\sum_{i=1}^{\kappa} \sum_{j=0}^{s(i)-1} \beta^{(x)}_{k,i,j} + \beta^{(u)}_{k,i,j} = \sigma.\label{eq:beta-sum}
\end{equation}
We then substitute \eqref{eq:zeta_func} with the following inequalities:
\begin{align} \label{eq:zeta_component}
 &\zeta_{k,i,j}^{(x)}(t)+ \beta^{(x)}_{k,i,j}\geq 0,\\\nonumber
 &\zeta_{k,i,j}^{(u)}(t)+ \beta^{(u)}_{k,i,j}\geq 0, \\\nonumber 
  &i = 1,\ldots,\kappa, \\\nonumber
  &j = 0,\ldots,s(i)-1,\forall t \in [t_k,t_{k+1}].
\end{align}

To simplify the notation, we will drop the subscript $k$ for the remainder of this section. The transformation of the constraints is justified by the following proposition.
\begin{proposition} \label{prop1}
There exist a set of $\{\beta^{(x)}_{i,j}, \beta^{(u)}_{i,j}\}$ such that \eqref{eq:beta-sum} and \eqref{eq:zeta_component} hold if and only if inequality \eqref{eq:zeta_func} holds. 
\end{proposition}
See appendix \ref{apdix:zeta_proof} for a proof. 

\subsection{CBF Lower Bound for Set Invariance}
\label{sec:CBFlowerbound}
As briefly anticipated in the previous section, the constraints in \eqref{eq:zeta_component} need to hold for every time instant in a given interval, resulting in an infinite number of constraints. To include such constraints in the MIQP formulation, we need to drop the dependency on $t$ while maintaining linearity in terms of $x$ and $u$. We perform one additional transformation by defining new variables that capture lower bounds (over time) of the expressions in \eqref{eq:zeta_component}:
\begin{align}\label{eq:zetaLowerBound}
  \zeta^{(x)}_{k,i,j,\min}=\min_{t\in[t_k,t_{k+1}]}  \zeta_{k,i,j}^{(x)}(t),\\ \nonumber \zeta^{(u)}_{k,i,j,\min}=\min_{t\in[t_k,t_{k+1}]}  \zeta_{k,i,j}^{(u)}(t),\\ \nonumber
  i\in\{1,\ldots,\kappa\}, j\in\{0,\ldots,s(i)-1\}.\nonumber
\end{align}
Then, \eqref{eq:zeta_component} can be exactly replaced by
\begin{multline} \label{eq:zeta_component_bound}
 \zeta_{k,i,j,\min}^{(x)}(t)+ \beta^{(x)}_{k,i,j}\geq 0,\;  
 \zeta_{k,i,j,\min}^{(u)}(t)+ \beta^{(u)}_{k,i,j}\geq 0, \\ 
  i\in\{1,\ldots,\kappa\}, j\in\{0,\ldots,s(i)-1\}.
\end{multline}
There is a finite number of such constraints, as they do not depend on continuous time anymore. We incorporate them into our MIP formulation in two steps.

The first step is to use the Big-$M$ encoding method to remove all the terms that are either monotonically increasing or bounded below by zero, and so they cannot be active at the current solution. 
We define sets of binary variables $z_{k,i,j}^{(x)}, z_{k,i,j}^{(u)} \in \{0,1\}$ for each one of the inequalities in \eqref{eq:zeta_component_bound}.
We then associate desired values of $\zeta_{k,i,j}^{(x)}(t)$ or $\zeta_{k,i,j}^{(u)}(t)$ according to the following rules:
\begin{equation*}
    z_{k,i,j}^{(x)} =
    \begin{cases}
      0, & c_{k,i,j}^{(x)T}x[t_k] \geq 0 \wedge \lambda_i \geq 0\\
      0, & c_{k,i,j}^{(x)T}x[t_k] \geq 0 \wedge \lambda_i \leq 0 \wedge \sigma \geq 0\\
      1, & \text{otherwise}
    \end{cases}
 \end{equation*}
 
 \begin{equation*}
    z_{k,i,j}^{(u)} =
    \begin{cases}
      0, & c_{k,i,j}^{(u)T}u[t_k] \geq 0 \wedge \lambda_i \geq 0\\
      0, & c_{k,i,j}^{(u)T}u[t_k] \geq 0 \wedge \lambda_i \leq 0 \wedge \sigma \geq 0\\
      1, & \text{otherwise}
    \end{cases}
  \end{equation*}
  These rules are motivated by the fact that when $z_{k,i,j}=0$, the corresponding inequality in \eqref{eq:zeta_component} is automatically satisfied, and hence it can be ignored. The rules are transformed into mixed-integer linear constraints using the big-$M$ method. 

For example, if we want to enforce $c_{k,i,j}^{(x)T}x[t_k] \geq 0 \iff z_{k,i,j}^{(x)} = 0$ and $c_{k,i,j}^{(u)T}u[t_k] \geq 0 \iff z_{k,i,j}^{(u)} = 0$ , the following mixed integer encodings are used:
 \begin{align*}
     &c_{k,i,j}^{(x)T}x[t_k] \leq M (1-z_{k,i,j}^{(x)}), \\ 
     &-c_{k,i,j}^{(x)T}x[t_k] \leq Mz_{k,i,j}^{(x)}, \\
     &c_{k,i,j}^{(u)T}u[t_k] \leq M (1-z_{k,i,j}^{(u)}), \\ 
     &-c_{k,i,j}^{(u)T}u[t_k] \leq Mz_{k,i,j}^{(u)}, \\
 \end{align*}
For $\lambda_i \geq 0 \iff z_{k,i,j}^{(x,u)} = 0$, we have the following
 \begin{align*}
   \lambda_i \leq M (1-z_{k,i,j}^{(x,u)}), \\ 
   -\lambda_i \leq M z_{k,i,j}^{(x,u)},
 \end{align*}
where $M$ is a sufficiently large number. 

For all terms such that $z_{k,i,j}^{(x,u)} = 1$, we need to ensure $\zeta_{k,\min}^{(x)}(x[t_k],\tau)$ and $\zeta_{k,\min}^{(u)}(u[t_k],\tau)$ are positive. Consider the CBF lower bound \eqref{eq:zetaLowerBound}, for $\forall t \in [t_k,t_{k+1}]$. The idea is that $\zeta_k(t)$ converges to a value monotonically when $j=0$ (simple eigenvalue) and has a minimum stationary point when $j \geq 1$ (Jordan block of dimension greater than one). Thanks to their simple forms, however, we can compute such lower bounds analytically in different cases as follows:
\begin{table}[b]
\centering
\begin{tabular}{|c|c|c|c|}
\hline
$\lambda_i$ & $j$ &
$\zeta_{k,\min}^{(x)}(x[t_k],\tau)$ & $\zeta_{k,\min}^{(u)}(u[t_k],\tau)$ \\
\hline
 $\geq 0$ & $\geq 1$ & $\zeta_{k,i,j}^{(x)}(\tau)$
 & $ \zeta_{k,i,j}^{(u)}(\tau)$\\
\hline
 $> 0$ & $=0$ &
$c_{k,i,j}^{(x)T} x[t_k] + \sigma + \beta_{k}^{(x)}$
 & $c_{k,i,j}^{(u)T} u[t_k] + \beta_{k}^{(u)}$\\
\hline
 $<0$ & $\geq 1$ &
 $\zeta_{k,i,j}^{(x)}(-\frac{j}{\lambda_i})$
& $\zeta_{k,i,j}^{(u)}(-\frac{j}{\lambda_i})$\\
\hline
 $<0$ & $=0$ &
$c_{k,i,j}^{(x)T} x[t_k] + \sigma + \beta_{k}^{(x)}$ 
 & $c_{k,i,j}^{(u)T} u[t_k] + \beta_{k}^{(u)}$\\
\hline
\end{tabular}
\label{tab1}
\caption{CBF Lower Bound with $c_{k,i,j}^{(x)T} x[t_k]\leq 0,c_{k,i,j}^{(u)T} u[t_k]\leq 0$}\label{tb:STL_CBF_lowerbound}
\end{table}

Note that the minimum values $\zeta_{k,\min}$ are linear in the optimization variables $x[t_k]$, $u[t_k]$. Therefore, these lead to linear constraints in our optimization problem.

\subsection{CBF Lower Bound for Always Operator}
\label{sec:CBFSTLencoding}

The forward invariance property from the CBF can be carried over to ensure STL satisfaction in continuous time. In short, we would like to enforce CBF constraints on all subformulae containing $\textbf{G}$ (always)
temporal operator, such that the continuous state trajectory satisfies the subformulae. More specifically, we first use the mixed-integer method from \ref{sec:STL-mixed-integer} to ensure the trajectory satisfies the formula at sampling instants $t_k, k=1,...,N$. Let us assume that we want to satisfy $\phi = \textbf{G}_{[t_1,t_2]} 
h_1 \geq 0$ where $h_1 := x_2(t) - 3$. Based on the integer encoding method above, assuming the system starts at $t=0$ and $z_{\phi}[t=0] = 1$, we have
\begin{align*}
    &z_{\phi}[t] \leq z_{h_1}[t_1],\\
    &z_{\phi}[t] \leq z_{h_1}[t_2],\\
    &z_{\phi}[t] \geq -1+z_{h_1}[t_1]+z_{h_1}[t_2].
\end{align*}

The formulation above ensures $z_{h_1}[t_1] = z_{h_1}[t_2] =1$, which implies $x_2[t_1] \geq 3 $ and $x_2[t_2] \geq 3 $. However, we cannot draw a conclusion in between $[t_1,t_2]$. 

To overcome this issue, we propose the following method: given a formula of the form $\varphi = G_{[a,b]} h(x) \geq 0 $ for some affine predicate $h(x)$, we can directly define the CBF constraint using the predicate. The idea is to ensure the state trajectory will stay within the set defined by predicate $h(x), \forall t \in [a,b]$.

\subsection{Predicate Lower Bound for Finite Time Reachability}\label{sec:predicate lowerbound}
Let us recall the linear predicate $h(x)$ and the corresponding set $C = \{x \in{\mathbb{R}^n}|h(x(t))  \geq 0 \}$. Given a predicate with eventually operator i.e., $\textbf{F}_{[t_k,t_f]} h(x) \geq 0$. Formally, given the system \eqref{continousSystem} and initial state $x(t_k) \not\in C$, find a control $u$, such that there exists a $t$ with $x(t) \in C, t_k \leq t \leq t_f$. To ensure finite time reachability, we denote a lower bound of $h(x(t))$ with respect to time, such that
\begin{equation}
\underline{h}(t) \leq h(x(t)), \forall t_k \leq t \leq t_f.
\end{equation}
Note it is trivial to see the following implication: $\underline{h}(t)\geq0 \implies h(x(t))\geq 0, \forall t_k \leq t \leq t_f$. The main idea is to find the lower bound $\underline{h}(t)$ that linearly depends on decision variables $x$ and $u$, such that we directly enforce $\underline{h}(t)\geq0$ as a constraint in the mixed-integer program. In contrast to the CBF lower bounds, we take a less conservative approach for finding $\underline{h}(t)$ such that some positive contributing terms i.e.,$c_{k,i,j}^{(x)T} x[t_k] e^{\lambda_i t} t^j,c_{k,i,j}^{(u)T} u[t_k] e^{\lambda_i t} t^j$ are not removed from the constraint. 

\subsubsection{Lipsthiz Constant Approach}
Given the initial time $t_k$, we can obtain the lower bound $\underline{h}(t)$ using descent lemma \cite{bertsekas1999nonlinear}:
\begin{align}\label{eq:h_lipschitz_lb}
\underline{h}(t) = h(t_k)+(t-t_k)\dot{h}(t_k)-(t-t_k)^2\frac{L}{2}\leq h(t),
\end{align}
for $t \geq t_0$ and Lipshtiz constant $L := \max \ddot{h}$; see \cite{bertsekas1999nonlinear} for a proof.

Let us recall the linear predicate $h(t) = \nu^{T} x(t) + \gamma$. With an abuse of notation, we can further expand it into
\begin{align*}
h(t)=\sigma + \sum_{i=1}^{\kappa} \sum_{j=0}^{s(i)-1} (c_{k,i,j}^{(x)T} x[t_k] e^{\lambda_i t} t^j+{c_{k,i,j}^{(u)T}}u[t_k] e^{\lambda_i t}t^j),
\end{align*}
which has the same form of the continuous-time CBF \eqref{eq:zeta_func}. However, the coefficients $c_{k,i,j}^{(x)}, c_{k,i,j}^{(u)}, \sigma$ are different since they depend on the predicate directly, instead of the CBF inequality constraint \eqref{eq:hocbf}. For the simplicity of the notation, we denote $x_k:=x[t_k],u_k:=u[t_k]$. Assuming $\lambda_i = 0, \forall i$ and $j=0,1,2$, we have 
 \begin{align*}
 &h(t) = c_{k,2}^{(x)T} x_k t^2+c_{k,2}^{(u)T} u_k t^2+c_{k,1}^{(x)T} x_k t^1+c_{k,1}^{(u)T} u_k t^1\\
 &+c_{k,0}^{(x)T} x_k+c_{k,0}^{(u)T} u_k + \sigma\\
 &\dot{h}(t) = c_{k,2}^{(x)T} x_k t+c_{k,2}^{(u)T} u_k t+c_{k,1}^{(x)T} x_k +c_{k,1}^{(u)T} u_k\\
 &\ddot{h}(t) = c_{k,2}^{(x)T} x_k+c_{k,2}^{(u)T} u_k
 \end{align*}
 To find Lipshtiz constant, we could utilize state bound $x_{max}$ and control bound $u_{max}$ to obtain $L = \max \ddot{h} = c_{k,2}^{(x)T} x_{max}+c_{k,2}^{(u)T} u_{max}$. Note the $\underline{h}(t)$ is linearly dependent on both $x$ and $u$, which we can later be used as a MIP constraint. For system with higher order, the rest of the terms ($\dot{h}(t),\ddot{h}(t)$) for \eqref{eq:h_lipschitz_lb} are expressed in appendix \ref{apdix:predicate_boud_lip}.
\begin{remark}
The Lipshtiz constant approach assumes that there exists a state bound and a control bound for the system. In addition, a large $x_{max}$ and $u_{max}$ could introduce conservatism for $\underline{h}(t)$.
\end{remark}

\subsubsection{Mixed-integer Approach}
We introduce another approach to obtain the lower bound $\underline{h}(t)$, which removes the dependency of state and control bounds. The lower bounds can be obtained from the mixed-integer encoding, according to Table \ref{tb:STL_Predicate_lowerbound}. The proofs can be found in Appendix \ref{apdix:main}.
\subsection{Predicate Lower Bound for Eventually Operator}\label{sec:predicate_lb_eventually}
Similar to the approach from section \ref{sec:CBFSTLencoding}, we can encode
the lower bounds as parts of the MIP constraints. Given an STL specification $\phi = \textbf{F}_{[a,b]} h(x(t)) \geq 0$ and  predicate lower bound $\underline{h}(t)$, we have
\begin{align}
    \underline{h}(x(t_k)) \geq 0 \implies x(t) \models \phi, \forall k=1,...,N.
\end{align}

\begin{table*}[ht]
\centering
\scalebox{0.80}{
\begin{tabular}{|c|c|c|c|c|c|}
\hline
$c_{k,i,j}^{(x)T} x[t]$&$c_{k,i,j}^{(u)T} u[t_k]$&$\lambda_i$&$j$&\textbf{Lower Bound} \\
\hline
$< 0$&$<0$&$\geq 0$&$\geq1$&  $\underline{h}_{k}^{(x)}(x[t_k]) = \frac{c_{k,i,j}^{(x)T} x[t_k] e^{\lambda \tau} \tau^j}{\tau}t $\\
 &&&&$\underline{h}_{k}^{(u)}(u[t_k]) = \frac{c_{k,i,j}^{(u)T} u[t_k] e^{\lambda \tau}\tau^j}{\tau}t $\\
\hline
 $<0$&$<0$&$\geq0$&$=0$& $\underline{h}_{k}^{(x)}(x[t_k]) = \frac{c_{k,i,j}^{(x)T} x[t_k] e^{\lambda \tau} \tau^j}{\tau}t  + c_{k,i,j}^{(x)T} x[t_k]$\\
&&&& $\underline{h}_{k}^{(u)}(x[t_k]) = \frac{c_{k,i,j}^{(u)T} u[t_k] e^{\lambda \tau} \tau^j}{\tau}t  + c_{k,i,j}^{(u)T} u[t_k]$\\
\hline
$<0$&$<0$&$\leq0$&$\geq0$& $\underline{h}_{k}^{(x)}(x[t_k]) = c_{k,i,j}^{(x)T} t^j-\lambda c_{k,i,j}^{(x)T}x[t_k]t^{(j+1)}+\frac{\lambda}{2}^2c_{k,i,j}^{(x)T}x[t_k]t^{j+2}$\\
 &&&& $\underline{h}_{k}^{(u)}(u[t_k]) = c_{k,i,j}^{(u)T} t^j-\lambda c_{k,i,j}^{(u)T}u[t_k]t^{(j+1)}+\frac{\lambda}{2}^2c_{k,i,j}^{(u)T}u[t_k]t^{j+2}$\\
 \hline
 $>0$&$>0$&$\leq0$&$\geq0$&  $\underline{h}_{k}^{(x)}(x[t_k]) = c_{k,i,j}^{(x)T} x[t_k]\lambda t^{j+1}+c_{k,i,j}^{(x)T}x[t_k]t^j$\\
&&&&  $\underline{h}_{k}^{(u)}(u[t_k]) = c_{k,i,j}^{(u)T} u[t_k]\lambda t^{j+1}+c_{k,i,j}^{(u)T}u[t_k]t^j $\\
\hline
$>0$&$>0$&$>0$&$=0$&  $\underline{h}_{k}^{(x)}(x[t_k]) = c_{k,i,j}^{(x)T} \lambda (\lambda-1)x[t_k] t^2 +c_{k,i,j}^{(x)T}x[t_k]t+c_{k,i,j}^{(x)T}x[t_k] $\\
&&&&  $\underline{h}_{k}^{(u)}(u[t_k]) = c_{k,i,j}^{(u)T} \lambda (\lambda-1)u[t_k] t^2 +c_{k,i,j}^{(u)T}u[t_k]t+c_{k,i,j}^{(u)T}u[t_k] $\\
\hline
$>0$&$>0$&$\geq 0$&$=1$&  $\underline{h}_{k}^{(x)}(x[t_k]) = c_{k,i,j}^{(x)T} x[t_k] e^{\lambda t} t^j $\\
&&&&  $\underline{h}_{k}^{(u)}(u[t_k]) = c_{k,i,j}^{(u)T} u[t_k]e^{\lambda t} t^j $\\
\hline
 $>0$&$>0$&$>0$&$=2$&  $\underline{h}_{k}^{(x)}(x[t_k],\tau) = c_{k,i,j}^{(x)T} \lambda j(j-1)x[t_k] t^2 $\\
&&&&  $\underline{h}_{k}^{(u)}(u[t_k],\tau) = c_{k,i,j}^{(u)T} \lambda j(j-1)u[t_k] t^2 $\\
\hline
\end{tabular}}
\label{tab3}
\caption{Predicate Lower Bound}\label{tb:STL_Predicate_lowerbound}
\end{table*}

\subsection{Optimization Problem}
\label{sec:miqp}
To solve Problem \ref{problemFormulation}, we formulate the following MIP: 
\begin{equation} 
\begin{aligned}
& \underset{\textbf{u},\textbf{x}}{\text{min}}
& &  J(u[t]) \\
& \text{s.t.}
& & x[t_{k+1}] = A_k x[t_k] + B_k u[t_k],\\
& & & x(t) 	\models \varphi, \\
& & & z_{\varphi}, z_{k,i,j}^{(x)}, z_{k,i,j}^{(u)}\in \{0,1\}, \\
& & & u_{l} \leq u_k \leq u_{u},\\
& & & k = 0,...,N-1 \\
& & & t \in [0,t_f], i=[1,...,\kappa],j=[1,...,s(i)].
\end{aligned}
\label{eq:optimizationProblem}
\end{equation}
The $N$ is the total number of controller updates and $t_f$ is the horizon of the formula $\varphi$. The decision variables for the MIP are $x[t_k]$ and $u[t_k]$ that are evaluated active time instants $t_{active}$. The $u_l$ and $u_u$ are the lower bound and upper control bounds, respectively. To ensure $x(t) \models \varphi$, we enforce the mixed integer constraints that are defined in \ref{sec:STL-mixed-integer} and \ref{sec:CBFlowerbound}. $A_k$ and $B_k$ are defined using the discretization method in Section \ref{sec:dynamical-system}. The cost function $J(u[t])$ can be selected either using quadratic cost i.e., $J(u[t])=u[t]^Tu[t]$ or $L1$ norm cost i.e., $J(u[t])=|u[t]|$.

\subsection{Nonuniform Control Updates}
In this formulation, both the number of control updates $N$ and update time instants $t_{active}$ need to be predetermined, which also effects the feasibility of \ref{eq:optimizationProblem}. We propose an heuristic approach to determine these parameters. First, we initialize a minimum number of control update $N_{min}$ based on the number of unique time instants based on time bounds from each predicates and decide the specific time instants, called active time instants,in which controls will be applied. For example, with $\varphi = \textbf{F}_{[0,2.1]} x_1 \geq 2 \wedge \textbf{G}_{[0,3.5]} x_2 \leq 5$, we have $N = 3$ with active time bounds $\{0,2.1\}$ for $\textbf{F}_{[0,2.1]} x_1 \geq 2$ and time bounds $\{0,3.5\}$ for $\textbf{G}_{[0,3.5]} x_2 \leq 5$. Therefore, we initialize active time instants $t_{active} = \{0,2.1,3.5\}$. 

We add all the initial constraints into \eqref{eq:optimizationProblem} and try to obtain a feasible solution. If the problem is not feasible, we perform bisection by adding additional update time instant in between all existing $t_{active}$ and solve again, the process continues until we obtain the solution or reach a pre-defined maximum iterations.

\begin{remark}
We can promote sparsity of the output control sequence by using $L1$ norm cost function in \eqref{eq:optimizationProblem}, which reduces the required control updates in actual implementation.  
\end{remark}

\begin{algorithm}[h] \label{adapative-algorithm}
\caption{Continuous STL Motion Planner}
\textbf{Input:} STL Formula $\varphi$\\
\textbf{Output:} $\{x[t_k]\},\{u[t_k]\},k=1,\dots,N$ ,
\begin{algorithmic}[1]
\State Initialize active instants $t_{active}$ from $\varphi$
\State Initialize control updates $N$
\State Initialize MIP  \eqref{eq:optimizationProblem}
\For {$N \leq N_{max}$}
    \State Solve for MIP 
    \If {MIP == feasible}
        \State \textbf{return} $\{x[t_k]\},\{u[t_k]\},k=1,\dots,N$ 
    \EndIf
    \State $t_{active} = bisection(t_{active})$
    \State $N=2N-1$
\EndFor
\end{algorithmic}

\end{algorithm}

\section{Results}
\subsection{Example 1: Continuous-time Eventually Operator}
In this example, we encode the $\textbf{F}$ operator with continuous predicate bound. Let us consider a one-dimensional double integrator system:
\begin{align} \label{eq:doubleIntSys}
\left[\begin{matrix}\dot{x_1}(t) \\ \dot{x_2}(t) \end{matrix} \right]= \left[ \begin{matrix} 0&1\\ 0&0 \end{matrix} \right] \left[\begin{matrix} x_1(t) \\ x_2(t) \end{matrix} \right]+ \left[ \begin{matrix} 0\\ 1 \end{matrix} \right] u,
\end{align}
where $x_1$ is the position and $x_2$ is the velocity. We would like the system to satisfy the following specification:
\begin{align*}
\varphi_1 := \textbf{F}_{[0s,1.0s]} (x_1(t) \geq 3 ) \wedge \textbf{F}_{[2.0s,4.5s]} (x_1(t) \leq -2) \end{align*}

\begin{figure*}[htbp]
\begin{minipage}[t]{0.32\linewidth}
    \includegraphics[width=\linewidth]{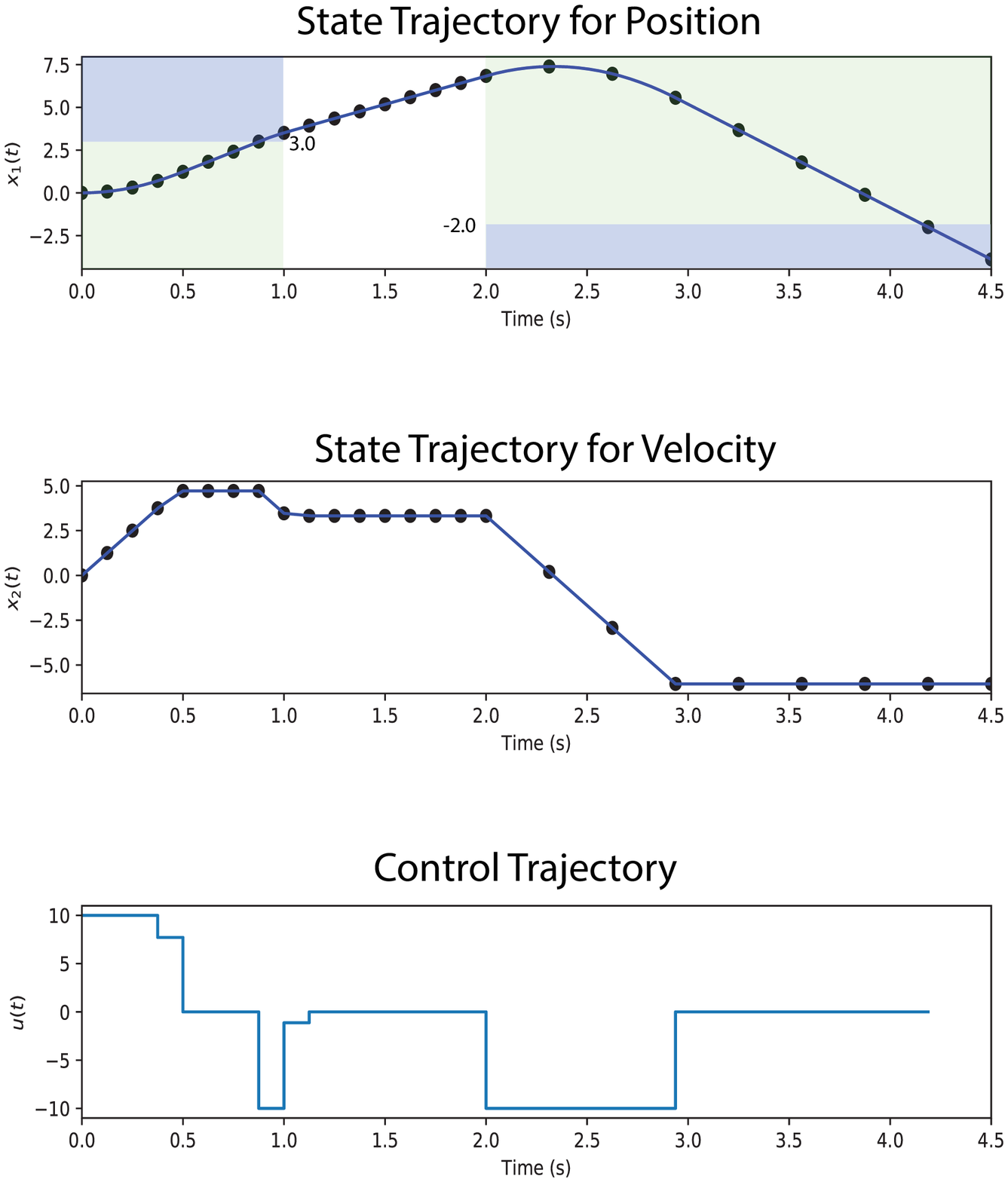}
\end{minipage}%
    \hfill%
\begin{minipage}[t]{0.32\linewidth}
    \includegraphics[width=\linewidth]{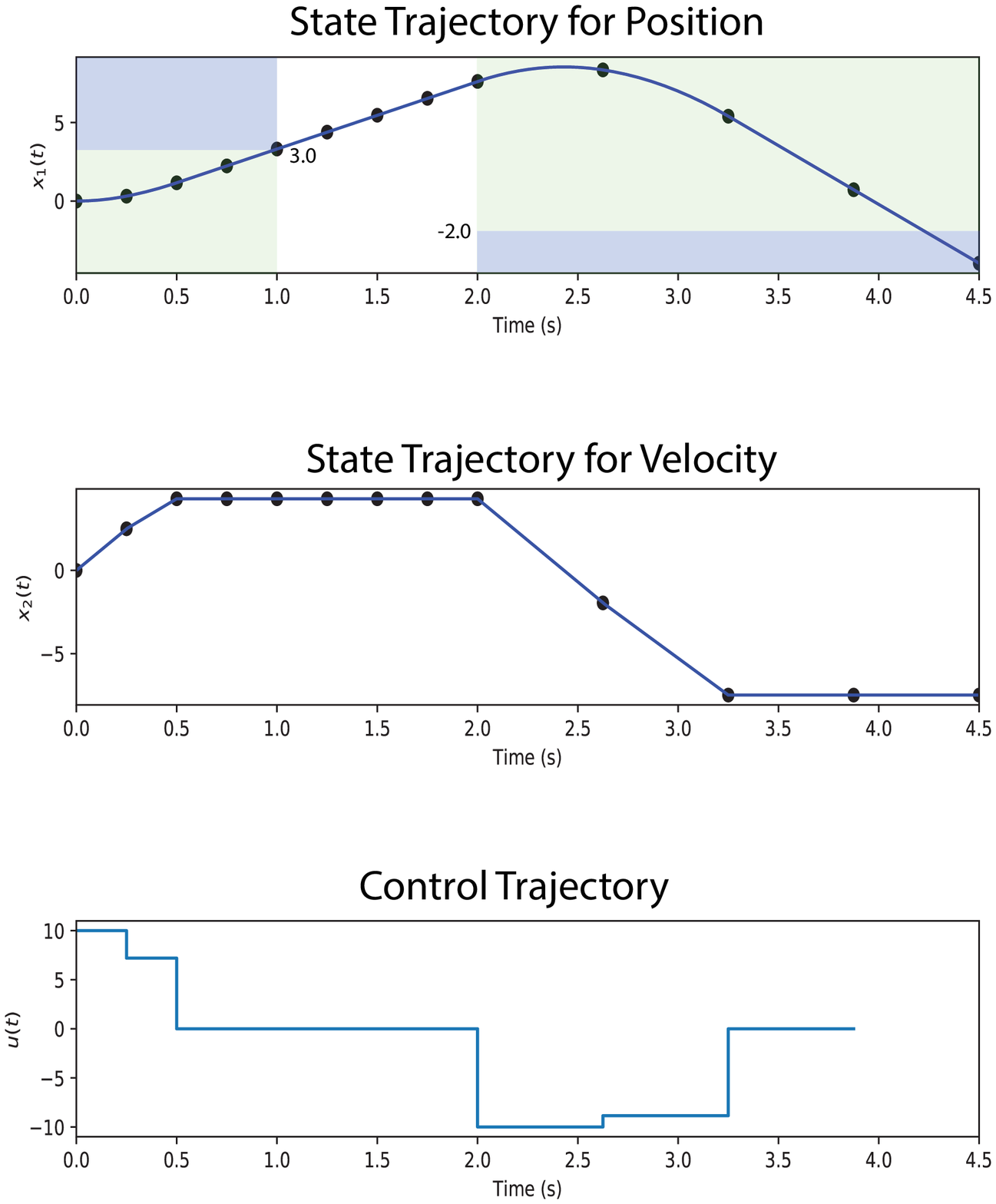}
\end{minipage}%
    \hfill%
\begin{minipage}[t]{0.34\linewidth}
    \includegraphics[width=\linewidth]{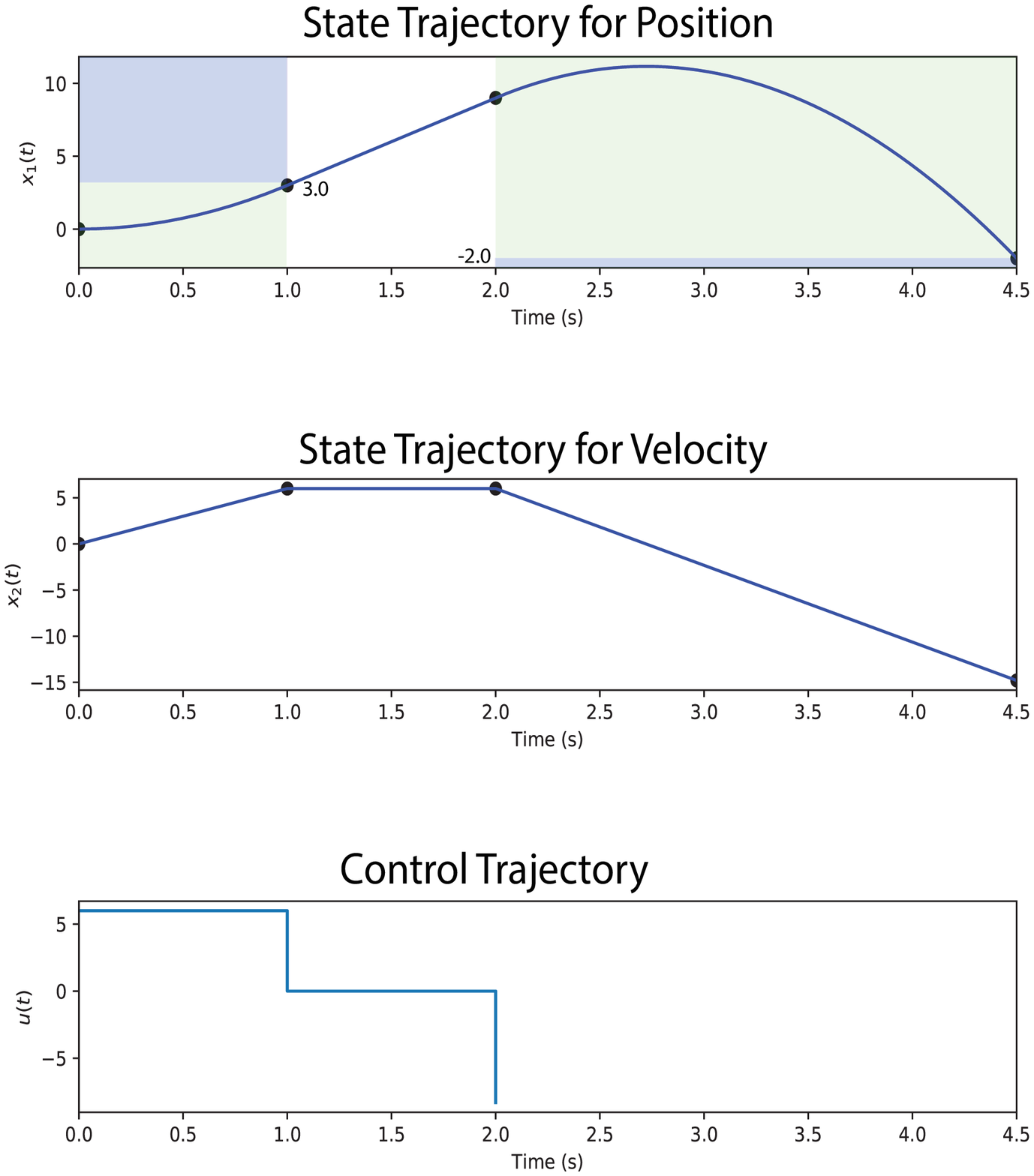}
\end{minipage} 
\caption{Example 2: Discrete $\textbf{F}$ (Left), Continuous $\textbf{F}$ using Lipshtiz Constant (Center), Continuous $\textbf{F}$ using mixed-integer encoding (Right)}
\label{fig:continuous-eventually-example}
\end{figure*}

We can interpret $\varphi_2$ as a finite-time reachability problem with discrete control updates. The position has to be greater or equal to 3 in between time $[0,1.0]$ and less or equal to -2 in between time $[2.0,4.5]$. The initial active time instants for this example are $t_{active}=\{0, 1.0, 2.0, 4.5\}$. The formulated problem is solved in 0.026s with a single iteration. The initial state $x_(t_0)=[0,0]^T$ with a total horizon $t_f=4.5$. We set the control bounds to be $[u_l, u_u]=[-10,10]$. The result is shown in Figure \ref{fig:continuous-eventually-example}.

Note, unlike discrete time $\textbf{F}$, where arbitrary number of active time instants have to be placed in time bounds $[0,1.0]$ and $[2.0,4.5]$, our continuous predicate bounds are only active at $t=0$ and $t=2.0$. The output trajectory $x_1(t)$ still satisfy predicates $\textbf{F}_{[0s,1.0s]} (x_1(t) \geq 3 )$ and $\textbf{F}_{[2.0s,4.5s]} (x_1(t) \leq -2)$. This approach greatly reduce the number of integers and constraints, without the need of specifying update instants.  

\subsection{Example 2: Continuous-time STL Motion Planning}
In this example, we demonstrate STL motion planning with both continuous Eventually $F$ and Always $G$ encoding method. Consider a two-dimensional double-integrator system:
\begin{align} \label{eq:2ddoubleIntSys}
\left[\begin{matrix}\dot{x_1}(t) \\ \dot{x_2}(t) \\ \dot{x_3}(t) \\ \dot{x_4}(t)  \end{matrix} \right]= \left[ \begin{matrix} 0&1&0&0\\ 0&0&0&0\\0&0&0&1\\0&0&0&0 \end{matrix} \right] \left[\begin{matrix} x_1(t) \\ x_2(t)\\ x_3(t)\\ x_4(t) \end{matrix} \right]+ \left[ \begin{matrix} 0&0\\ 1&0 \\0&0 \\0&1 \end{matrix} \right] \left[ \begin{matrix} u_1 \\ u_2\end{matrix} \right],
\end{align}

where $x_1$, $x_3$ are positions and $x_2$, $x_4$ are velocities.
We would like to satisfy the following STL formula with horizon $t_f = 1s$:
\begin{align} \label{eq:STL_example_spec}
\varphi_2 := &\textbf{F}_{[0.1s,0.6s]} (x_1(t) \leq -0.5 \wedge x_3(t) \geq 0.5) \nonumber\\ &\wedge \textbf{F}_{[0.7s,1s]} (x_1(t) \geq 1 \wedge x_3(t) \geq 1)\\ \nonumber &\wedge \textbf{G}_{[0s,1s]} (x_1(t)\geq 0 \vee x_3(t) \geq 0),\\ \nonumber &t\in {[0,t_f]}. \end{align} 
The initial state $x(t_0)= [1,0,-0.5,0]^T$ with CBF constants $[k_1, k_2]=[30,30]$. We set the control bound as $[u_{l}, u_{u}]=[-40,40]$ for both $u_1$ and $u_2$. The MIP is solved in 0.06 seconds. 
\begin{figure}[ht]
    \centering
    \includegraphics[width=\linewidth]{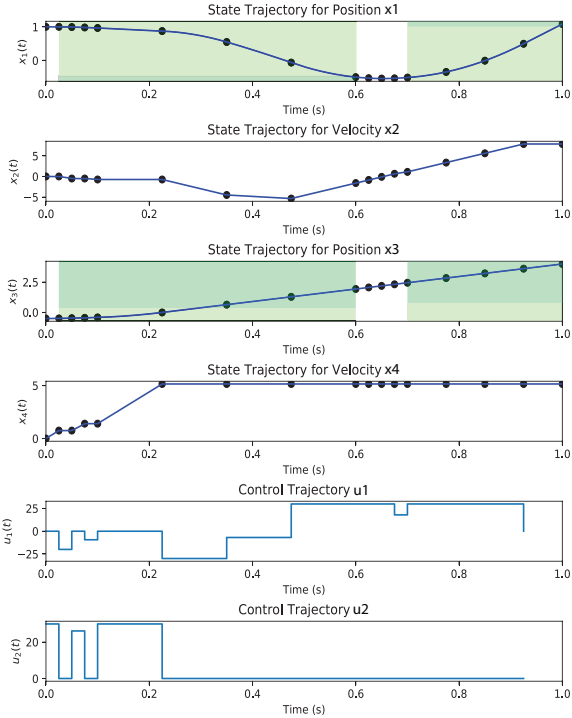}
    \caption{Example 2: State trajectories over time}
    \label{Example 2: State Trajectories over time}
\end{figure}
\begin{figure}[ht]
    \centering
    \includegraphics[width=1.0\linewidth]{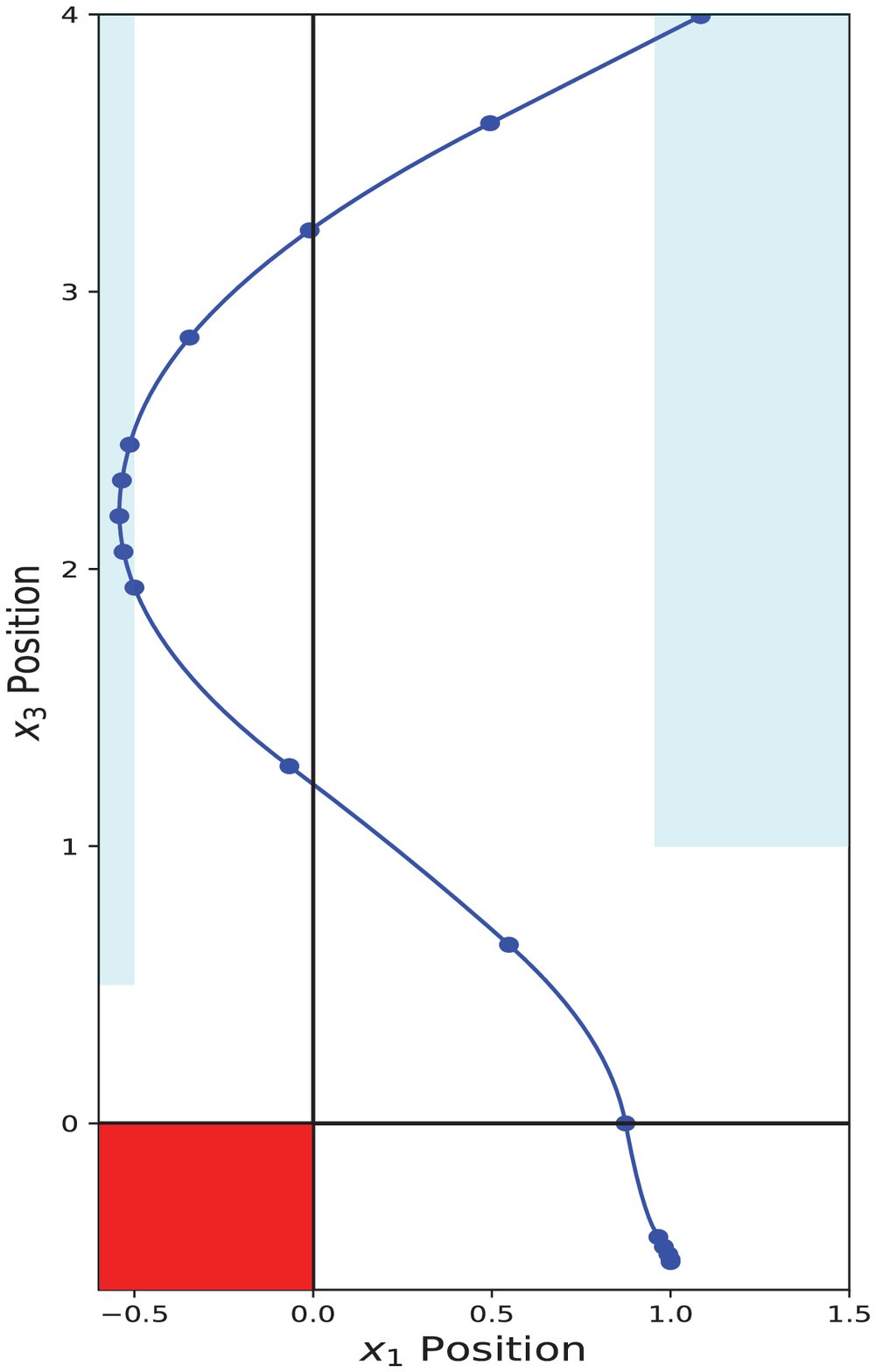}
    \caption{Example 2: Workspace Trajectory}
    \label{fig:my_label}
\end{figure}

\section{Conclusions}
In this paper, we proposed an optimization-motion planner under continuous-time STL. The motion planner automatically determines the appropriate sampling and control updates instants. The algorithm is validated in double integrator systems. For future works, we would like to implement MPC and perform experiments on quad-copters to simulate real-world scenarios. 

\appendix
\section{Appendix}
\label{apdix:main}
\subsection{Proof for Proposition \ref{prop1}}\label{apdix:zeta_proof}
\begin{proof}
  To prove that \eqref{eq:zeta_component} implies \eqref{eq:zeta_func}, we can simply sum all the inequalities in \eqref{eq:zeta_component} over $i$ and $j$, and then simplify the summation of the $\beta$'s using \eqref{eq:beta-sum}.
  
To prove that \eqref{eq:zeta_func} implies \eqref{eq:zeta_component}, we first define the ``excess'' quantity
\begin{equation}\label{eq:excess}
\delta = \sum_{i=1}^{\kappa} \sum \limits_{j=0}^{s(i)-1} \bigl( \zeta_{i,j}^{(x)}(t) + \zeta_{i,j}^{(u)}(t) \bigr),
\end{equation}
and then construct the $\beta$'s by splitting $\delta$ and $\sigma$ evenly as follows:
\begin{align*}
    \beta^{(x)}_{i,j} =& - \zeta_{i,j}^{(x)}(t) + \frac{\delta+\sigma}{2 \sum_{i=1}^n (s(i)-1)}, \\ 
    \beta^{(u)}_{i,j} =& - \zeta_{i,j}^{(u)}(t) + \frac{\delta+\sigma}{2 \sum_{i=1}^n (s(i)-1)}.\\
  &i\in\{1,\ldots,\kappa\}, j\in\{0,\ldots,s(i)-1\},\nonumber
\end{align*}

We first verify that these $\beta$'s satisfy the summation constraint \eqref{eq:beta-sum}:
\begin{multline}
    \sum_{i=1}^{\kappa} \sum_{j=0}^{s(i)-1} \beta^{(x)}_{i,j} + \beta^{(u)}_{i,j} = -\sum_{i=1}^{\kappa} \sum_{j=0}^{s(i)-1} ( \zeta_{i,j}^{(x)}(t) +  \zeta_{i,j}^{(u)}(t)) \\+ 2 \sum_{i=1}^{\kappa} \sum_{j=0}^{s(i)-1} \frac{\delta+\sigma}{2 \sum_{i=1}^{\kappa} (s(i)-1)} = \sigma
\end{multline}

To show that the constructed $\beta$'s also satisfy \eqref{eq:zeta_component}, first notice that by substituting \eqref{eq:excess} into \eqref{eq:zeta_func}, we have $\delta+\sigma\geq 0$; then we have
\begin{equation}
  \zeta_{ij}^{(x)}(t)+ \beta^{(x)}_{i,j}=\frac{\delta+\sigma}{2 \sum_{i=1}^n (s(i)-1)}\geq 0,
\end{equation}
for all $i,j$, and with an analogous expression for $\zeta_{ij}^{(u)}(t)$, $\beta^{(u)}_{i,j}$. This completes the proof.

\end{proof}

\subsection{Predicate Bound with Lipshtiz Approach}\label{apdix:predicate_boud_lip}
 For the simplicity of the notation, we denote $x_k:=x[t_k],u_k:=u[t_k]$. The terms for \eqref{eq:h_lipschitz_lb} are expressed in the rest of the section. In the case where system has higher orders, and assuming $\lambda_i \geq 2, j \geq 2$, we have
\begin{align*}
    &\dot{h}(t_k) = \sum_{i=1}^{\kappa} \sum_{j=0}^{s(i)-1} c_{k,i,j}^{(x)T} \lambda_i x_k e^{(\lambda_i-1) t} t^j+c_{k,i,j}^{(x)T} x_k e^{\lambda_i t}j t^{(j-1)}\\&
    +{c_{k,i,j}^{(u)T}}u_k \lambda_i e^{(\lambda_i-1) t}t^j
    +{c_{k,i,j}^{(u)T}}u_k e^{\lambda_i t}jt^{(j-1)},\nonumber
\end{align*}
\begin{align*}
 &\ddot{h}(t_k) = \sum_{i=1}^{\kappa} \sum_{j=0}^{s(i)-1} c_{k,i,j}^{(x)T} \lambda_i  (\lambda_i-1)x_k e^{(\lambda_i-2) t} t^j\\&+c_{k,i,j}^{(x)T} x_k  \lambda_i e^{(\lambda_i-1) t}j t^{(j-1)}+c_{k,i,j}^{(x)T} \lambda_i x_k e^{(\lambda_i-1)t}jt^{(j-1)}\\&+c_{k,i,j}^{(x)T} x_k e^{\lambda_it}j(j-1)t^{(j-2)}+c_{k,i,j}^{(u)T} \lambda_i  (\lambda_i-1)u_k e^{(\lambda_i-2) t} t^j\\&+c_{k,i,j}^{(u)T} u_k  \lambda_i e^{(\lambda_i-1) t}j t^{(j-1)}+c_{k,i,j}^{(u)T} \lambda_i u_k e^{(\lambda_i-1)t}jt^{(j-1)}\\&+c_{k,i,j}^{(u)T} u_k e^{\lambda_it}j(j-1)t^{(j-2)}
\end{align*}

\subsection{Proof for Predicate Lower Bounds}
In this section, we only prove for predicate lower bound for a component that contains $x_k$. The same proof can be directly carried over for a component that has $u_k$.

Suppose $h(t)=c_{k,i,j}^{(x)T} x_k e^{\lambda t}t^j$, with  $c_{k,i,j}^{(x)T} x_k \leq 0$, $\lambda \geq 0$ and $j \geq 1$, we can directly obtain the lower bound $\underline{h}(t)= \frac{c_{k,i,j}^{(x)T} x_k e^{\lambda \tau} \tau^j}{\tau}t$ for $t_0 \leq t \leq t_0 + \tau$ using Jenseng's inequality \cite{boyd2004convex}. The same technique can be carried over for $c_{k,i,j}^{(x)T} x_k \leq 0$, $\lambda \geq 0$ and $j = 0$. The lower bound is then $\underline{h}(t)= \frac{c_{k,i,j}^{(x)T} x_k e^{\lambda \tau} \tau^j}{\tau}t+c_{k,i,j}^{(x)T}x_k$

 For $c_{k,i,j}^{(x)T} x_k \leq 0$, $\lambda \leq 0$ and $j \geq 0$, the lower bound is $\underline{h}(t)=c_{k,i,j}^{(x)T} t^j-\lambda c_{k,i,j}^{(x)T}x_kt^{(j+1)}+\frac{\lambda}{2}^2c_{k,i,j}^{(x)T}x_kt^{j+2}$. 
\begin{proof}
 Without the loss of generality, we assume $\lambda = -1$ and $c_{k,i,j}^{(x)T}x_k = -1$, next we can expand out $e^{-t}$ as
 \begin{align*}
 e^{-t} &= 1-t+\frac{1}{2}t^2-\frac{1}{3!}t^3+\dotsc\\
 &\leq 1-t+\frac{1}{2}t^2\\
 \end{align*}
given $h(t)=-e^{-t}t^j$, we have
\begin{align*}
h(t) &= -(1-t+\frac{1}{2}t^2-\frac{1}{3!}t^3+\dotsc)t^j\\
&\geq -(1-t+\frac{1}{2}t^2)t^j = \underline{h}(t)
\end{align*}
In general, we have 
\begin{align*}
h(t) &\geq c_{k,i,j}^{(x)T}x_k t^j-\lambda c_{k,i,j}^{(x)T}x_k t^{(j+1)}+\frac{\lambda}{2}^2c_{k,i,j}^{(x)T}x_k t^{j+2}\\
&=\underline{h}(t).
\end{align*}
\end{proof}
 For $c_{k,i,j}^{(x)T} x_k > 0$, $\lambda \leq 0$ and $j \geq 0$, the lower bound is $\underline{h}(t)=c_{k,i,j}^{(x)T}x_k t^j$.
 \begin{proof}
 Given $c_{k,i,j}^{(x)T} x_k > 0$, $\lambda \leq 0$ and $j \geq 0$, we have
 \begin{align*}
 h(t)&=c_{k,i,j}^{(x)T} x_k e^{\lambda t}t^j\\
 &= c_{k,i,j}^{(x)T} x_k (1+\lambda t+\frac{\lambda^2t^2}{2}+\dotsc)t^j\\
 &\geq c_{k,i,j}^{(x)T}x_k (1+\lambda t)t^j = \underline{h}(t)
 \end{align*}
 \end{proof}
 
 For $c_{k,i,j}^{(x)T} x_k \geq 0$, $\lambda \geq 0$ and $j=0$, the lower bound $\underline{h}(t)=c_{k,i,j}^{(x)T} \lambda (\lambda-1)x[t_k] t^2 +c_{k,i,j}^{(x)T}x[t_k]t+c_{k,i,j}^{(x)T}x[t_k]$
 
 \begin{proof}
For $c_{k,i,j}^{(x)T} x_k \geq 0$, $\lambda \geq 0$ and $j=0$, we have
 \begin{align*}
 h(t)&=c_{k,i,j}^{(x)T} x_k e^{\lambda t}\\
 &= c_{k,i,j}^{(x)T} x_k (1+\lambda t+\frac{\lambda^2t^2}{2}+\dotsc)\\
 &\geq c_{k,i,j}^{(x)T}x_k (1+\lambda t+\frac{\lambda^2t^2}{2}) = \underline{h}(t)
 \end{align*}
 \end{proof}
 
  \begin{proof}
For $c_{k,i,j}^{(x)T} x_k \geq 0$, $\lambda \geq 0$ and $j=1$, we take the time derivative
 \begin{align*}
 \dot{h}(t)=c_{k,i,j}^{(x)T} x_k e^{\lambda t}+c_{k,i,j}^{(x)T} x_k\lambda  e^{\lambda t}t,
 \end{align*}
 \end{proof}
 By taking the Taylor expansion at $t_k$, we have
  \begin{align*}
 h(t) &\geq c_{k,i,j}^{(x)T} x_k e^{\lambda t_k}+(c_{k,i,j}^{(x)T} x_k e^{\lambda t_k}+c_{k,i,j}^{(x)T} x_k\lambda  e^{\lambda t_k}t_k)t\\&=\underline{h}(t)
 \end{align*}
 Assuming $t_k=0$, we have
 \begin{equation*}
 \underline{h}(t)=c_{k,i,j}^{(x)T} x_k t    
 \end{equation*}
 
 The same method carries over for $c_{k,i,j}^{(x)T} x_k \geq 0$, $\lambda \geq 0$ and $j=2$, assuming $t_k=0$, we have $\underline{h}(t) = c_{k,i,j}^{(x)T} \lambda j(j-1)x_k t^2$.

\bibliographystyle{IEEEtran}        
\bibliography{IEEEabrv,mybib}           

\end{document}